# How Could Equality and Data Protection Law Shape AI Fairness for People with Disabilities?

Reuben Binns

University of Oxford, Department of Computer Science, reuben.binns@cs.ox.ac.uk

Reuben Kirkham

Monash University, Department of Human-Centred Computing, reuben.kirkham@monash.edu

This article examines the concept of 'AI fairness' for people with disabilities from the perspective of data protection and equality law. This examination demonstrates that there is a need for a distinctive approach to AI fairness that is fundamentally different to that used for other protected characteristics, due to the different ways in which discrimination and data protection law applies in respect of Disability. We articulate this new agenda for AI fairness for people with disabilities, explaining how combining data protection and equality law creates new opportunities for disabled people's organisations and assistive technology researchers alike to shape the use of AI, as well as to challenge potential harmful uses.

## 1 INTRODUCTION

'AI fairness' is one name used for a growing field of interdisciplinary research in which issues of fair treatment, discrimination, and justice are considered in relation to artificial intelligence systems, particularly those based on machine learning. Much of this work focuses on protected characteristics such as race or gender, and more recently some scholars have focused on specific dimensions of AI fairness for people with disabilities (PWD). [1]

The development of AI fairness is both motivated, supported, and constrained by relevant laws, and AI fairness in the context of PWD is no exception. We observe that there are significant differences in how the law interacts with AI in relation to PWD as compared to other protected characteristics. As Morris argues: "*lawmakers must grapple with whether and how rules regarding the use of AI systems might change depending on an end user's disability status*" (Morris 2019). There are also multiple different areas of law and regulation which impact on AI fairness for PWD, including disability human rights law, equality law, and privacy and data protection law, which can also differ significantly between jurisdictions. The question of how lawmakers, assistive technology designers and HCI practitioners should grapple with the law around AI systems and PWD is therefore a complex and contextually-dependent one.

---

[1] We acknowledge the difference between 'identify first' and 'person first' terminology and recognise that there are a variety of perspectives on this. In this article, we use person first terminology throughout, to be consistent with both the SIGACCESS accessible writing guide (https://www.sigaccess.org/welcome-to-sigaccess/resources/accessible-writing-guide/, archived at https://perma.cc/6FWZ-Z9Y9), as well as the legislation, including the UN Convention on the Rights of Persons with Disabilities (A/RES/61/106) (the main human rights treaty concerning disability), which follows this approach throughout.

This article narrows this broad question down to focus on two major bodies of law which have significant implications for the development of AI fairness in the context of PWD, namely: data protection law and equality law. Data protection law protects individuals' fundamental rights in relation to the processing of their personal data (including but not limited to privacy).[2] Equality law protects individuals from direct and indirect discrimination on the grounds of protected characteristics, such as race, sex, and disability status. It also offers special protections for PWD, most notably the duty to provide reasonable accommodations or adjustments.

Whilst data protection and equality law are by no means the only relevant areas of law for AI fairness and disability, they are arguably two of the most important in the EU.[3] Fundamentally, they address some of the central dimensions at issue in the context of AI and PWD, and are both carefully crafted (and implemented) to align with more general obligations, such as human rights law (especially the European Convention on Human Rights (ECHR)).

For the purposes of this article, we focus on these laws as they are enacted in the United Kingdom. While this does narrow the scope of our analysis somewhat, focusing on a specific jurisdiction will allow for a more in-depth analysis. UK law is particularly worth examining because (i) it is one of the more longstanding legal systems to have disability discrimination legislation, (ii) the UK system is particularly developed and operates within a single equality law regime (principally the Equality Act (2010)) and (iii) unlike many other jurisdictions (e.g. the USA), the UK has a general regulation covering personal data - the GDPR - which is the main driving force behind recent advances in data protection law, and now a globally recognised standard. There is also the advantage in that the UK system aligns with the ECHR, which is the world's longest standing human rights framework. Thus by grounding our argument in UK law, we offer a forward-looking exemplar which can be built upon by the assistive technology and AI fairness communities for years to come. Furthermore, because both of these UK legal instruments are partly or wholly derived from European Union law, our analysis will likely raise the same or similar implications for AI fairness and PWD in the European Union member states more broadly.[4] While the UK's exit from the European Union does open up the possibility for divergence in future, at the time of writing the UK statutes are just as well aligned with EU law than the remaining EU member states.

---

[2] For readers unfamiliar with the distinction, the following examples may help: there are aspects of privacy not covered by data protection, for instance where privacy is breached without data being processed, such as in voyeurism; there are aspects of data protection not covered by privacy, such as those relating to correcting innacurate data or the prohibition on solely automated decision making. For more on the distinction between data protection and privacy, see Gellert & Gutwirth 2013.

[3] Other significant areas include consumer law (Lazar 2019)

[4] At the same time, there are also Commonwealth nations with a similar system to the UK (e.g. Australia, Canada, India) who also are increasingly adopting GDPR-style provisions. To give an example, the Australian state of Victoria has the Charter of Human Rights and Responsibilities Act 2006 which is similar to the UK Human Rights Act (1998), whilst also having a single equality Act which borrows frequently from UK provisions, and is in the process of increasingly tightening its data protection law, especially for public sector organisations.

Data protection and equality law have many potential implications in the context of AI fairness for PWD. These include obligations imposed on technology designers working in the context of public services or private companies deploying AI systems, which may support efforts towards fairness for PWD, and restrictions on the use of data about PWD which may promote or inhibit efforts towards fairness. There are also ways in which these two areas of law may be in tension when applied to certain methods proposed to promote fairness for PWD.[5]

In posing and offering answers to these questions, our aim is to highlight ways in which the efforts of AI Fairness for PWD can be supported by existing equality and data protection law in the EU; as well as how those laws may in some cases present barriers or challenges to such efforts. We argue that, due to the unique ways in which disability is treated in equality law, it is important to have a separate body of principles and approaches towards AI fairness when designing either for PWD (e.g. the development of assistive technologies), or creating systems to be used by the public at large by way of universal design (which includes PWD). At the same time, we offer a positive new agenda for assistive technology researchers and practitioners (as well as disabled persons' organisations), who may be able to take advantage of opportunities provided by legislation to more effectively advance the rights of persons with disabilities.

To anticipate what follows, **Section 2** explains the substantive law, **Section 3** explains how AI can be unfair to PWD (including existing technical approaches for helping to mitigate this risk), then **Section 4** provides an analysis of how both data protection and equality law intersect with the AI fairness agenda for PWD, before **Section 5** explains how PWD can be protected from inappropriate AI systems, and presents a wider agenda for both assistive technologies researchers and disability advocates, harnessing the opportunities created by fully considering data protection law and equality law.

**2. EQUALITY AND DATA PROTECTION LAW**

This section introduces some fundamental aspects of equality and data protection law.[6] Given the inherent difficulties in interdisciplinary work like this, the introduction is necessarily non-comprehensive; our perspectives here are shaped by our respective experiences working on the boundary between computer science and law.[7] Our aim in this section is to cover salient aspects to the extent needed to draw out their implications for AI fairness and PWD. We address Equality Law first, as this is the conceptual framework of non-discrimination (and thus acts as the closest corollary of 'AI fairness' in the legal domain), before examining how this aligns to data protection law (where non-discrimination can be factored into the way that personal data is processed in computational systems).

---

[5] For instance, where data protection law restricts the use of disability status as an input to an automated decision about an individual, but equality law might in practice require it (see Section 4.2.2).
[6] For more in-depth introductions, see McColgan 2014 in respect of discrimination law and Coppel 2020 for information rights law, including data protection.
[7] To provide readers with a better understanding of the author's respective backgrounds and how they relate to the interdisciplinary challenges this paper addresses, we provide the following brief 'positionality statements' which explain our relationship with legal research (we thank an anonymous reviewer for this suggestion). One author (Binns) draws heavily in this paper from experience as an applied public interest technologist working for a legal regulator applying data protection law to AI, as well as holding an interdisciplinary PhD between computer science and law. In addition to research in HCI with legal implications (e.g. Binns et al 2018; Van Kleek et al 2018; Agrawal et al 2021), Binns has published legal scholarship in technology law journals (e.g. Veale, Binns & Edwards 2018; Veale, Binns & Ausloos 2018; Binns 2019; Binns & Bietti 2020). The other author (Kirkham) has worked on the boundary of interactive technologies and the legal process, including empirical legal research (Kirkham, 2018), the design of systems that address legal problems (e.g. Watson, Kirkham, & Kharrufa 2020) and how discrimination law can be complementary to the design of AT (e.g., Kirkham 2015; Kirkham & Greenhalgh 2015; Kirkham 2020). Kirkham is interested in the pragmatics of AT design and how this can be supported by properly considering legislation, so that PWD can fully benefit from AT in the real world. Our joint aim is to provide a basic overview of these two areas of law and their broad application to issues of AI fairness for PWD. While both authors are well-versed in the relevant legal scholarship, and have practical experience in the relevant legal issues, we do not attempt to undertake conclusive legal analysis of any particular AI systems.

We have provided a relatively detailed analysis of the relevant law, due to the level of its complexity and the need for precision, as well as to enable adaptation of this work in face of future legal changes. In many respects, this is similar to a qualitative thematic analysis, applied to the underlying legal material (see for example Kirkham, 2020). To make this account more accessible to non-lawyers, and to enable a quicker read by someone who wishes to grasp the core principles as they stand, we have related some of this material to end-notes that appear at the end of this document.

2.1. **Equality law**

2.1.1. The general operation of Equality Legislation.

Most work on AI fairness has so far focused on race or gender, rather than disability. This means that it is necessary to understand general Equality Legislation and how its core concepts closely align to the concern of AI fairness in general (including for people *without* disabilities), so this can be contrasted with the different approach required for people with disabilities. In many countries, equality law can be complicated, especially where different pieces of legislation address the needs of different groups, yet the UK simplified this by creating a single equality act[8], the Equality Act (2010), which provides protections for nine 'protected characteristics', namely: "*age; disability; gender reassignment; marriage and civil partnership; pregnancy and maternity; race; religion or belief; sex; sexual orientation*".[9] The main effect of the Equality Act (2010) is to prohibit discrimination in certain areas of life: (i) Services and Public Functions; (ii) Premises, (iii) Work (and Employment), (iv) Education and (v) Associations.[10] It only applies to these areas, rather than all activities.[11]

For most protected characteristics there are two main types of discrimination to be concerned with:

***Direct discrimination:*** This is the classic form of discrimination that can take place and is what most people would intuitively think discrimination is. Direct discrimination occurs when "*because of a protected characteristic*", a person or organisation treats another person "*less favourably than [they] treats or would treat others.*"[12] This might be said to be 'classic' discrimination, but has some qualifications that are not straightforward. The relevant treatment has to be l*ess favourable*, so treatment that is differential does not automatically (but in practice often does) amount to less favourable treatment and thus discrimination[13]. Likewise, it is possible to treat groups the same but still less favourably, for example by segregating them from one another.[14] The test for direct discrimination is *objective*, and thus excludes consideration of motive as a defence[15], meaning that well intentioned acts can thus amount to direct discrimination.[16] Indeed, direct discrimination can even be done *subconciously*.[17] There is normally no possibility of justifying direct discrimination:[18] *less favourable treatment* is enough to found a successful claim, as long as it was *because of* the protected characteristic.

---

[8] See Hepple 2010.
[9] S.4 of the Equality Act (2010). More specific definitions can be found in s.4-12.
[10] These are covered by different parts of the Act (Parts 3 to 7).
[11] For instance, it would not be unlawful for someone not to provide an access ramp for a wheelchair user at their private home, to attend a private dinner party unrelated to work (or another area covered by the Act). In these circumstances, there would be no duty to make reasonable adjustments *at all,* as it would fall outside the ambit of the Act.
[12] s.13 of the Equality Act (2010).
[13] Smith v Safeway plc [1996] IRLR 456

*<u>Indirect discrimination:</u>* This concerns matters that apply to everyone, but can have disproportionate effects on one group: this form of discrimination is perhaps the most relevant and challenging with respect to AI fairness. Formally, indirect discrimination concerns a "*provision, criterion or practice*" (often called a 'PCP') that is "*discriminatory in **relation** to a relevant protected characteristic*"[19] There are some important observations that should be made about this form of discrimination. First, a PCP is a broad concept and can include a wide range of concerns, from contractual requirements (e.g. an employment contract requiring inflexible working hours[20]) to uncodified workplace rules (such as keeping one's face visible[21]). Unlike direct discrimination, there is no need for *every* person with the relevant characteristic to be negatively impacted by the PCP in question; it is enough that it is *more likely* to impact them.[22] Second, it is not required for there to be "*explanation of the reasons why a particular PCP puts one group at a disadvantage when compared with others […] It is enough that it does.*"[23] (a point that is particularly pertinent to AI systems, see (Grimmelmann & Westreich, 2017)). Third, it is possible (but not *neccessary*) to show indirect discrimination with statistics, comparing one group to another. Finally, unlike direct discrimination, indirect discrimination generally provides for the possibility of justification, provided there is a "*legitimate aim*" and the PCP is "*proportionate*" in respect of that aim.

**<u>Exemptions:</u>** In addition to the fact that Equality Legislation only applies to specified areas of activity, there are a few exemptions to Equality Legislation that apply in special circumstances (e.g. religious organisations) and also where an organisation is engaging in positive action (this is particularly limited in the sphere of employment, but limited more generally, too): positive action is not allowed to become positive discrimination. Further, an organisation may not 'contract out' from the Act, with any contract that purports to do so being invalid in law.[24]

---

[14] Chief Inspector of Education, Children's Services and Skills v The Interim Executive Board of Al-Hijrah School (Rev 2) [2017] EWCA Civ 142

[15] Birmingham City Council v Equal Opportunities Commission [1989] AC 1155, and James v Eastleigh Borough Council [1990] 2 AC 751

[16] See for example Moyhing v Barts and London NHS Trust [2006] IRLR 860.

[17] Consider for instance the case of Geller & Anor v Yeshurun Hebrew Congregation [2016] UKEAT 0190_15_2303.

[18] With the exception of narrow cases related to the protected characteristic of 'age', where direct discrimination can be justified in certain circumstances.

[19] s.19 of the Equality Act (2010). Technically, this is a four part test, all of which have to be satisfied for someone to successfully claim discrimination: (i) the PCP applies to other people than those with the characteristic in question (if it does not, then it is direct discrimination); (ii) the PCP "*puts, or would put, persons with whom [a complainant] shares the characteristic at a particular disadvantage when compared with persons [that the claimant] does not share it with*", (iii) the PCP puts the *complainant themselves* at a disadvantage and (iv) the person (or organisation) applying the PCP "*cannot show it to be a proportionate means of achieving a legitimate aim*".
This does not apply to 'pregnancy or maternity' as a protected characteristic, however the other eight are included.

[20] London Underground Limited v Edwards (No 2) [1998] IRLR 364.

[21] Azmi v Kirklees Metropolitan Borough Council [2007] IRLR 434

[22] For someone to bring a claim, they *themselves* have to be subject to negative treatment, see Footnote 19 above.

[23] Essop & Ors v Home Office (UK Border Agency) [2017] UKSC 27 at [24].

[24] Part 10 of the Equality Act (2010) provides that it is not possible to avoid the obligations posed by that Act by contracting out (or to enforce a contract which "*constitutes, promotes or provides for [unlawfully discriminatory] treatment*"), meaning that an

**Motivation of compliance:** An important consideration in understanding the practical effects of discrimination law is to understand the mechanisms that motivate compliance with it. For public authorities, the duty to avoid discrimination is generally an active one: for instance, s.149 (the Public Sector Equality Duty (PSED)) requires a public sector organisation to have 'due regard' to the need to eliminate discrimination and advance equal opportunities. However, given the nature of existing legal tests and that discriminating by accident (or subconsciously) is no defense, any organisation (including a private one) would be unwise not to have fully evaluated the potential implications of their own practices. What's more, there are a range of civil penalties - both individuals (as "*employee[s]*" and/or "*agents*") and organisations can be sued for discrimination. A further concern is that civil liability attaches to someone who "*instructs*", "*causes*" or "*induces*"[25] , whether successfully or not, someone to breach most aspects of the Act[26]: this is of particular concern to people who hold out as relevant experts, as the provision of inaccurate information in such a capacity could amount to a breach of the act. There are also criminal offences, punishable by a fine.[27]

At the same time, there are also soft (non-legal) provisions, such as professional codes that explicitly expect compliance with the principles of non-discrimination law within the course of one's work as a professional. This includes documents such as the ACM Code of Ethics[28] or the Bar Standards Board handbook[29]. Nevertheless, despite these insistent legislative demands, there is a limitation with equality law, in that it is often not fully implemented in practice, largely due to the complaint-based legal system, where individuals have to bring litigation before Courts or Tribunals to enforce their rights (whilst risking the costs and expenses of undergoing this process[30]).

---

understanding of potentially discriminatory implications are important to have in mind (and to have sufficient knowledge about) when forming any contractual relationship, lest that relationship becomes in part a nullity down the line (per s.142(1) and s144(1).). In respect of disability, this also extends to 'non-contractual' terms in the employment context (per .s.142(2) and s.142(3); s144(2) and s144(3).

[25] Per s.111 of the Act. Whilst this is constrained to 'basic contraventions', this is a broad term that encompasses the great majority of discrimination covered in the Act, including that discussed in this article. This includes the case where the (attempted or actual) inducement is indirect.

[26] Per s.111 of the Act. Whilst this is constrained to 'basic contraventions', this is a broad term that encompasses the great majority of discrimination covered in the Act, including that discussed in this article.

[27] See Part 8 of the Equality Act (2010). It is also an criminal offence for someone in the role of "*employer or principal*" to *"knowingly or recklessly [give] a statement which is false or misleading in a material respect"* to the effect of authorising unlawful discrimination (per s 110(3) and s.110(4)). More generally, it is also a criminal offence for someone to "*knowingly or recklessly make [a false or misleading] statement*" to the effect that any action that helps someone contravene the Equality Act (2010) is in fact lawful (per S.112. This is also limited to 'basic contraventions'). In other countries, such as Australia, there is the possibility of imprisonment in addition to the stigma of a criminal conviction. For instance, s.43 the Disability Discrimination Act 1992 (Cth) makes it an offence of "*to incite … ,assist or promote*" unlawful discrimination under said Act, with a penalty of (up to) "*six months imprisonment*" on conviction.

[28] Per 1.4 and 2.3 of the Code.

[29] The Code of Conduct provides that a barrister "must not discriminate unlawfully against any person".

[30] The risk of costs depend upon the forum: in Employment and Special Educational Needs, these jurisdictions are normally cost neutral (i.e. each party bears their own legal costs) as they are determined within administrative tribunals, whilst someone challenging goods or services would have to apply to the County Court, where cost-shifting usually takes place (asides in first-instances small claims).

2.1.2. *How and why is disability different?*

We now turn to analyze the legal approach for PWD and highlight how it is fundamentally distinctive to other protected characteristics. Under the UK Equality Act (2010), disability is an expansive concept. The formal test is that the person "*has a physical or mental impairment*" and "*the impairment has a substantial and long-term adverse effect on [that person's] ability to carry out normal day-to-day activities*"[31], with an impairment being long term if it has lasted more than 12 months, or is likely to last for either that period of time or the end of the person's life[32]. The definition of disability also includes conditions that can recur over time.[33] The result is that there are a broad range of impairments[34] that have been accepted as disabilities under the Act in practice, including physical impairments (e.g. a back condition that limits the ability to lift heavy objects at work[35], spinal muscular atrophy[36] or being a manual wheelchair user[37]), many medical conditions (e.g. cancer[38], a peanut allergy[39] or eczema[40]), vision impairments[41], hearing impairments[42], and a range of cognitive impairments (e.g. Dyslexia[43] and Learning Disabilities[44]) as well as psychological conditions (e.g. Depression[45], Amnesia[46] or Schizophrenia[47]). The result is that about 18% of the working-age population is considered in law to have a disability[48] however, perhaps more salient is the diversity of people (and impairments) protected from discrimination.

The nature and extent of discrimination against people with disabilities necessitates a legal system that provides additional protections and in particular, an additional type of discrimination, namely the failure to provide reasonable adjustments (or accommodations[49]). It means that there is a need for a separate consideration of how they could and should be protected from discrimination. There are also other important protections in the United Kingdom, namely the protection from 'discrimination arising from a disability'. We explain how each of these approaches operate in turn:

---

[31] s.6 of the Equality Act (2010). There are some qualifications found in Schedule 1 of the Act, including provision for regulations that further define disability.
[32] Paragraph 2 of Schedule 1.
[33] Paragraph 2 of Schedule 2.
[34] The UN Convention on the Rights of Persons with Disabilities provides a wider consideration of what constitutes a disability, with less of an emphasis on a medical model approach (see e.g. (Koyithara 2018) for a discussion on this).
[35] Banaszczyk v Booker Ltd (Disability Discrimination: Disability) [2016] UKEAT 0132_15_0102.
[36] Royal Bank of Scotland Group Plc v Allen [2009] EWCA Civ 1213
[37] Firstgroup Plc v Paulley [2014] EWCA Civ 1573
[38] Schedule 1 of the Equality Act (2010) at Paragraph 6.
[39] As found by the Employment Tribunal in Wheeldon v Marstons plc (ET/1313364/12, 9 May 2013)
[40] This was found in respect of a severe case by an Employment Tribunal in Glass v Promotion Line (ET/3203338/12, 24 Apr 2013).
[41] The Equality Act 2010 (Disability) Regulations 2010 provide that someone is disabled for the purposes of the act if they are "*certified as blind, severely sight impaired, sight impaired or partially sighted by a consultant ophthalmologist*".
[42] Cordell v Foreign and Commonwealth Office [2012] ICR 280. In respect of hearing impairments, we recognise that some people regard themselves as being Culturally Deaf, however the way they are protected by law is by having an "*impairment*" that limits their ability to process or access sound (which in turn can mean the provision of a Sign Language interpreter is required as a reasonable adjustment, where this is appropriate).
[43] *Paterson v Commissioner of Police of the Metropolis* [2007] UKEAT/0635/06/2307
[44] Dunham v Ashford Windows [2005] UKEAT 0915_04_1306
[45] As one example of many, see the Employment Tribunal case of Knopwood v Secretary of State for Work and Pensions (ET/2402722/15, 24 Nov 2015).
[46] Sobhi v Commissioner of the Police of the Metropolis UKEAT/0518/12/BA.
[47] Burdett v Aviva Employment Services Ltd UKEAT/0439/13/JOJ
[48] https://www.papworthtrust.org.uk/about-us/publications/papworth-trust-disability-facts-and-figures-2018.pdf (archived at: https://perma.cc/AYV9-6L9W)
[49] This latter phrasing is used in most countries outside of the UK, being found in the UN CRPD.

**Reasonable adjustments:** This provision is a distinctive right specifically for people with disabilities. The duty to make reasonable adjustments focuses on the individual circumstances of someone with a disability. More specifically, where "*a provision, criterion or practice*" or "*a physical feature*" (or a lack of an "*auxiliary aid*") "*puts a disabled person at a substantial disadvantage in relation to a relevant matter in comparison with persons who are not disabled*", the responsible organisation must "*take such steps as it is reasonable*" for it to take to "*avoid the disadvantage*" (or "*provide the auxiliary aid*").

This means in effect a two-part test: first the person with a disability needs to demonstrate a disadvantage (arising either from a PCP, 'physical feature' or lack of an 'auxiliary aid') and if so, then the organisation (*not* the disabled person[50]) is responsible for taking sufficient steps to remove said disadvantage, so long as those steps are legally "*reasonable*". The word 'reasonable' is inherently deceptive and more insistent in its demands than what a non-lawyer might intuitively expect, because it relates to an *objective* and "*purposive*"[51] standard (the purpose being the substantive inclusion of people with disabilities in wider society), not a subjective or intuitive one.[52] This 'objectivity' can be double-edged: from a positive perspective, Courts and Tribunals have required organizations to undertake expensive or difficult to implement reasonable adjustments (albeit not at unlimited expense[53]), thus recognising the extensive steps that can be sometimes needed to ensure equality of opportunities for PWD.[54] Furthermore, a proposed adjustment need not have a high likelihood of being successful in order to be legally reasonable: even adjustments that are unlikely to be successful can be reasonable, as long as there is "*just a prospect*" of success.[55] The flip side of the objective approach is that the organisation, not the PWD, can decide which approach to use to eliminate the disadvantage, with steps such as redeployment (against the will of the employee in question) having been held as being sufficient to discharge the duty[56]: moreover it is even possible for an organisation to comply with the duty by accident.[57]

Another important point is that for employment, the legal duty is only individual and *reactive* [58] to be liable, whilst for services and public functions there is an *anticipatory duty* to make reasonable adjustments for disabled people they have not met.[59] Where it applies, this anticipatory duty is also insistent in its demands and expectations.[60] However, with services, the duty to make reasonable adjustments only applies to the service itself, rather than to change the fundamental character of the service to be more suitable for the person with a disability (e.g. a restaurant is not required a takeaway service, even if that is the only way someone could access the food it

---

[50] Project Management Institute v Latif [2007] IRLR 579.
[51] Matthew Goodwin v Patent Office [1999] IRLR 4.
[52] See British Gas Services v McCaull 2001 IRLR 60 and Tarbuck v Sainsbury Supermarkets Ltd [2006] IRLR 664.
[53] See Cordell v Foreign and Commonwealth Office [2012] ICR 280 for an example of this.
[54] See for instance: (Kirkham 2015a, 2015b), for some potential advantages of this in the context of assistive technologies.
[55] Leeds Teaching Hospital NHS Trust v Foster (Disability Discrimination : Reasonable adjustments) [2011] EQLR 1075.
[56] Hay v Surrey County Council [2007] EWCA Civ 93.
[57] See British Gas Services v McCaull 2001 IRLR 60 at 42 and 43
[58] In particular, the employing organisation must have at the least, constructive knowledge of the relevant disability to be held liable for discrimination Gallop v Newport City Council [2013] EWCA Civ 1583.
[59] Roads v Central Trains Ltd. [2004] EWCA Civ 1541 and Finnigan v Chief Constable of Northumbria Police [2013] EWCA Civ 1191; FirstGroup Plc v Paulley [2017] UKSC 4.
[60] Examples include Commissioner of Police for the Metropolis v ZH [2013] EWCA Civ 69 and Royal Bank of Scotland v Allen [2009] EWCA Civ 1213.

provides[61]): for the purposes of this article, this likely means that an organisation who provides an *intrinsically* AI-based service is not going to be required by equality law to provide an alternative as a reasonable adjustment.[62]

**Direct discrimination is expressly permitted against people *without* disabilities:** A notable feature of s.13 (direct discrimination) of the Equality Act (2010) is that it is **not** direct discrimination to treat persons with a disability "*more favourably*". Indeed, the approach of reasonable adjustments necessarily involves sometimes treating people with disabilities more favourably. As was explained by the UK's highest court (at the relevant time), disability discrimination law: "*does not regard the differences between disabled people and others as irrelevant. It does not expect each to be treated in the same way. It expects reasonable adjustments to be made to cater for the special needs of disabled people. It necessarily entails an element of more favourable treatment.*"[63] This is distinct from other protected characteristics, which can be said to be 'symmetric' in most circumstances: for instance (barring an exemption), it is just as unlawful to discriminate on the basis of sex against a woman as it is a man, despite the historically asymmetric nature of sex discrimination.[64]

**Discrimination arising from a disability**: This is an important legal innovation that is distinctive to the United Kingdom and illustrates how far a progressive legal test for disability discrimination can go, in effect serving as a broader disability-specific form of discrimination that recognizes the unique circumstances that apply to disability. For someone to be discriminated against, they have to be treated "u*nfavourably because of something arising in consequence of [their] disability*" and the organisation or person engaged in that treatment "*cannot show that the treatment is a proportionate means of achieving a legitimate aim.*"[65] This test has important distinctions. First, the emphasis is on a "*something*", which is a wider term than a PCP; often making a claim easier to bring relative to either a reasonable adjustments claim, or indirect discrimination.[66] Second, the emphasis on "*something*" loosens the causal link required for discrimination to take place, removing the need for a comparator (unlike direct discrimination).[67] Third, the "*something*" can be loosely connected: it need not be the sole or main cause of the unfavourable treatment, with the influence simply having to be "more than trivial".[68] The result of the introduction of 'discrimination arising from disability' has therefore led to some curious cases being upheld on appeal: including the wheelchair user dismissed for racist remarks (his defence was he was angry due to a failure to make reasonable adjustments)[69] and the person suffering from a "*schizophrenic illness*" who committed sexual assaults and was dismissed on this basis[70]: these types of decisions at least implicitly give more weight to disability than other protected characteristics. However, the caveat is that this test does require *constructive knowledge* of the disability itself (unlike an anticipatory reasonable adjustment claim against a service provider or public authority, or a claim of indirect discrimination)[71]: nevertheless, this caveat does not apply to the fact that the specific "*something*" arises from the disability, just the existence of the "*disability*" itself.[72] From the point of view of AI, the practical effect of this test is that are a wider range of circumstances where the otherwise discriminatory use of an AI would need to be justified, potentially serving as a powerful tool, albeit one that presently only applies in the UK, but may also be implemented in other jurisdictions in the future.

---

[61] Edwards v Flamingo Land Ltd [2013] EWCA Civ 801
[62] If that system happened to directly discriminate against someone with a disability, this would be another matter.
[63] Archibald v. Fife Council [2004] UKHL 32 at [47]
[64] The question of whether this asymmetric approach should be extended to other protected characteristics is a live one in certain jurisdictions and can be authorised on a case-by-case basis (e.g. as in Victoria, Australia); however, our aim here is only to outline UK law as it currently stands.
[65] s.15(1). .
[66] See the discussion in General Dynamics Information Technology Ltd v Carranza [2015] IRLR 43 at [33] and [34]. However, there are cases where there is an exemption from s.15 but not s.19. One example is competence standards, wherein 53(7) provides that "*the application by a qualifications body of a competence standard to a disabled person is not disability discrimination unless it is discrimination by virtue of section 19."*
[67] See Land Registry v Houghton & Ors [2015] UKEAT 0149/14/1202 at [4] for a discussion of this deliberate legislative choice.
[68] Pnaiser v NHS England & Anor UKEAT/0137/15/LA at [31].
[69] Risby v London Borough of Waltham Forest [2016] UKEAT 0318/15/1803
[70] Burdett v Aviva Employment Services Ltd UKEAT/0439/13/JOJ

**Enforcement of Disability Discrimination Law:** The main issue with disability discrimination law is its weak enforcement, requiring people with disabilities (and all the disadvantages that can bring) to bring proceedings to enforce their rights. This means that despite the extent of the expectations offered by disability law, its actual implementation on the ground is somewhat limited. The (UK) Equality and Human Rights commission has observed that "*disabled people are still being treated as second-class citizens … the Equality Act 2010 has still not been implemented in full, …, life chances for disabled people remain very poor, and public attitudes to disabled people have changed very little*"[73], whilst the UN has found that many people in the UK with disabilities are subject to "*grave [and] systematic violations*" of their human rights, especially in relation to those who have particular support needs from the state.[74] From a statistical perspective, the wider conditions that PWD *on average* face in the UK are most troubling. To give some examples: a disability employment gap of nearly 30%[75]; reduced pay for those PWD who do have employment[76]; a markedly increased rate of poverty amongst families with one member who has a disability[77] (including being more than twice as likely to be in food poverty[78]), whilst men with mental health conditions have their life expectancy reduced by 20 years.[79] At the same time, the legislation has worked for some disabled people, just to a far more limited effect than is needed on the wider population of people with disabilities (Harwood, 2016). One of the major problems is *access to justice*: there is very limited availability of legal representation for most legal claims, whilst bringing a claim against a service provider risks being hit with the other side's legal costs.[80] What's more, Courts and Tribunals (including the judiciary) can have similarly problematic attitudes, with widespread failures in making reasonable adjustments having been recently reported.[81] As such,

---

[71] s.15(2)
[72] City of York Council v Grosset [2018] EWCA Civ 1105 at [51].
[73] https://www.equalityhumanrights.com/sites/default/files/being-disabled-in-britain.pdf (archived at: https://perma.cc/X56D-ECUN)
[74] UN Report: CRPD/C/15/R.2/Rev.1
[75] https://www.ons.gov.uk/peoplepopulationandcommunity/healthandsocialcare/disability/bulletins/disabilityandemploymentuk/2019 (archived at https://perma.cc/8RB3-NN4S)
[76] https://www.papworthtrust.org.uk/about-us/publications/papworth-trust-disability-facts-and-figures-2018.pdf (archived at: https://perma.cc/AYV9-6L9W)
[77] https://www.jrf.org.uk/data/poverty-rates-families-disabled-person (archived at: https://perma.cc/2WBE-G4KZ)
[78] https://www.equalityhumanrights.com/sites/default/files/being-disabled-in-britain.pdf (archived at: https://perma.cc/HQV2-LVT3)
[79] https://www.equalityhumanrights.com/sites/default/files/being-disabled-in-britain.pdf (archived at: https://perma.cc/HQV2-LVT3)
[80] See the evidence given to Parliament by a Mr Doug Paulley, a serial user of the Equality Act (2010) to bring disability discrimination claims: http://data.parliament.uk/writtenevidence/committeeevidence.svc/evidencedocument/women-and-equalities-committee/disability-and-the-built-environment/written/44155.pdf (archived at: https://perma.cc/C65T-K44F)
[81] See Galo v Bombardier Aerospace UK [2016] NICA 25, which found that an entire Tribunal in the United Kingdom had made no appropriate arrangements to ensure that reasonable adjustments were made by disabled litigants, or John Horan's account of the treatment of PWD by the judiciary (see Horan 2011 and Horan 2015)

disability discrimination law can be a powerful tool, but is only available in practice to a minority of PWD, and even then only with the use of much persistence and energies on their part.

**The United Convention on the Rights of Persons with Disabilities (the UN CRPD)**. Disability discrimination law has an international dimension to it, which is primarily articulated by way of the UN CRPD. At this point it is worth observing that the model, in terms of what the legislation outlined above requires, aligns closely with the core principles of the UN CRPD, even though the convention itself is not transposed into UK law. This is important because the UN CRPD is the key human rights treaty for PWD and thus a detailed legislative framework aligns with that Convention will be generally applicable elsewhere (in other words, our analysis is likely to hold true in most circumstances outside of the UK). However, the UN CRPD is at the same time widely not implemented in the UK - because the legislation is at best, irregularly enforced, and insufficient resources are provided to support the more vulnerable people with disabilities (e.g. those with social care needs, or those on disability benefits).[82] This is an odd position, where the UK law represents an effective model that we can use for our forward-looking analysis, yet presents a picture on the ground that sadly - in common with all countries - does not provide fully equal opportunities to people with disabilities.[83] In this paper we point to Data Protection Law as a potential means to overcome some of the limitations of discrimination law in the context of AI fairness, with a view towards enhancing the opportunities available for people with disabilities.

2.2. **Data protection law**

The main regulatory instrument for data protection law is the General Data Protection Regulation[84], an EU Regulation which has direct effect in EU member states (and in the UK, via the Data Protection Act 2018 which remains in place despite the UK leaving the EU).

As defined in European Union law, data protection law aims to protect fundamental rights and freedoms in relation to the processing of personal data (including privacy, but also other rights such as non-discrimination, including disability discrimination). As such, data protection may present opportunities to protect PWD from harms resulting from the processing of their personal data by an AI system. Personal data includes any information relating to an individual (known as the 'data subject') who can be identified directly (e.g. a name or government identifier) or indirectly (e.g. unique location data) (Article 4(1)). 'Processing' includes any operation on the data, including storing, analysing, deleting, combining, or sharing (Article 4(2)). An *organisation*[85] (whether public or private) which processes such data is a 'data controller' if they determine the purposes and means of the processing (Article 4(7), or a 'data processor' if they do so under instruction of another data controller (Article 4(8)).

---

[82] See for example the shadow report of the (UK) Equality and Human Rights Commission (Equality and Human Rights Commission, 2018), or the UN CRPD Committee's own report on the United Kingdom.
(CRPD/C/15/R.2/Rev.1)
[83] See generally the recent edited collection by Hisayo Katsui and Shuaib Chalklen (Katsui and Chalklen, 2020)
[84] Notably, the absence of data has been said to be one of the main obstacles of implementing the UN CRPD (see Mittler, 2015 for a discussion of this problem). At the same time, there have been long standing concerns in respect of disability, information privacy and security (see Lazar et al, 2017, which frames this as a human rights concern).
[84] GDPR: General Data Protection Regulation (EU) 2016/679
[85] There is an exemption where an individual is processing data for their own purposes, with processing by a "natural person in the course of a purely personal or household activity" being excluded from the ambit of the GDPR, per Article 2(c)

**Principles and lawful basis**

Data controllers must identify a specific purpose for processing personal data (Article 1(b)) and identify an appropriate lawful basis for that purpose, from the six available (Article 6(1)a-f). These include consent of, or **contract** with, the data subject, or a **legal obligation** or legitimate interest of the controller. The selected lawful basis must be appropriate to the context. For instance, if there is a power imbalance between the controller and the data subject (e.g. in employer-employee relations), consent would be inappropriate in so far as the data subject lacks the genuine freedom to grant or deny consent.[86] In relation to AI and PWD, an employer could not require employees with disabilities to consent to having their data processed by an AI system as a condition of employment; the choice must be genuinely voluntary and not subject to punitive consequences for denying consent. Furthermore, controllers must only process data that is adequate, relevant, and necessary for their purposes, ensure all such data is accurate, only store this data for as long as necessary and put in place appropriate measures to secure and protect it (Article 5(1)a-f). Data processed for one purpose cannot later be used for different incompatible purposes; for instance, an organization that collected disability data to assess whether their AI system was exhibiting errors for PWD, could not then use that same data for advertising purposes.

**Special category data**

Certain types of personal data are designated as 'special category' data (Article 9); these partly overlap with some of the protected characteristics in equality law, including race, religion, and disability (which is included within the definition of 'health' (Recital 35)). The category itself need not be explicitly encoded in the data; as any data processing 'revealing' the category can constitute special category data. In order to process special category data, in addition to having one of the six lawful bases mentioned above, controllers also need to fulfill one of ten further conditions. These include where the controller has the 'explicit consent' of the data subject, and for 'reasons of substantial public interest'.

Some of these further conditions are in turn further spelled out in national law. For instance, if special category data is being processed on the basis that it is for 'reasons of substantial public interest', there are 23 such reasons defined in Schedule 1 of the UK Data Protection Act 2018. They include 'equality of opportunity or treatment', and 'support for individuals with a particular disability or medical condition' (these two conditions are particularly pertinent to the analysis in Section 4).

**Solely automated decisions**

Where personal data processing is the sole basis for a decision which produces legal or 'similarly significant' effects on the data subject (such as being automatically denied a job opportunity or financial loan, or being offered a certain insurance premium), there are additional restrictions on data controllers (Article 22). Such decisions are prohibited except where the lawful basis of the processing is contract, legal obligation or explicit consent. Where such decisions are permitted, the data subject must have 'suitable safeguards' including the right to obtain human intervention, express their point of view, and contest the decision (and if the processing is based on a legal obligation, those safeguards must be set out in the law). For instance, a PWD has the right to contest the outcome of a solely automated, significant decision made about them by an AI system, to obtain human intervention, and to express their point of view (e.g. that they believe the system has failed them because of their disability).

---

[86] Guidelines 05/2020 on consent under Regulation 2016/679 Version 1.1

Furthermore, if a solely automated decision is based on special categories of personal data (such as disability), this also constrains the range of available conditions for processing special category data to just two: explicit consent and substantial public interest. As a result, there are only two ways a controller could justify using a system which uses special category data to make a solely automated decision which significantly affects an individual. First, where the controller has obtained the explicit consent of the data subject. And second, where a) the processing on which the decision is based is necessary for a contract or a legal obligation, and b) one of the substantial public interest conditions apply. This substantially constrains the circumstances under which it would be lawful to use an AI system to make a solely automated, significant decision on the basis of the data subject's disability status.

### **Accountability and Data Subject Rights**

Data controllers must also put in place measures to ensure they are accountable for their compliance, and document their processes so they are able to prove that they are compliant (Article 5(2)). They are required to implement the data protection principles 'by design and default' rather than merely as an afterthought (Article 25). They also need to conduct data protection impact assessments (DPIAs) for certain kinds of processing deemed 'high risk', which is likely to be the case for AI systems, especially where they have the potential to discriminate against PWD. A DPIA should include an assessment of the risks to a data subject's rights and freedoms, and the measures taken to address those risks; data subjects or their representatives should be consulted as part of a DPIA where appropriate.

In addition to these rules placed on data controllers, data subjects have a range of rights regarding the processing of their data, including: to be informed about various aspects of the processing of their personal data, to access a copy of their personal data by way of a subject access request), to rectify inaccurate data, to have data erased, to restrict and object to its processing, and to 'port' their personal data between services (Articles 13-21).

3. **AI FAIRNESS AND PWD**

A variety of research in recent decades has explored the legal, ethical and political dimensions of artificial intelligence, machine learning and various forms of algorithmic decision making. Much of this relates to questions of discrimination, justice, and equity, often encompassed by the term fairness. While we appreciate that there are important differences between and disagreements about these terms, for brevity (and following this special issue) we refer to this emerging sub-field collectively as 'AI fairness'.[87] We briefly outline this sub-field as a precursor to the legal implications. First, we discuss the ways in which AI systems may be 'unfair' to PWD; second, we outline some commonly proposed algorithmic methods of assessing and mitigating or 'fixing' unfairness. As we explain in the next section, equality and data protection law have the potential to be used to challenge these 'unfair' AI systems; but they also have important implications for the viability of certain algorithmic methods for assessing and mitigating such 'unfairness' which have been proposed in the AI fairness literature.

---

[87] See e.g. Bennett and Keyes (2019), who argue that concerns about AI *fairness* and PWD should be reoriented around questions of *justice*; similar positions have also been articulated outside the context of PWD, e.g. Greene et al (2019), Hoffman (2019), Binns (2018).

### 3.1. **How AI systems may be 'unfair' to PWD**

Several recent contributions to this field have set out a research agenda for AI fairness and PWD.[88] As identified in the call for papers for this special issue, this agenda can be mapped out in terms of the major ways that research on AI fairness and PWD intersect.[89] First, there are what we might call AI-driven assistive technologies (AT), "*AI systems designed specifically for accessibility scenarios (e.g., sign language recognition)".* Second, there are concerns raised by "*AI tools whose use disproportionately impacts people with disabilities*". This covers AI which may be about managing disability in various ways, such as tools for determining benefits eligibility, or other state / organisational decision making, but is not designed **for** disabled people as such. Finally, there are concerns raised by "*general-purpose AI systems (e.g., ensuring that mainstream tools are accessible to people with disabilities)".*

In the latter case, AI systems are not specifically designed as assistive technology to help PWD, but rather are intended to be deployed to the general population of people both with and without disabilities. As with all technology, if this is not inclusive by design, it is likely that it will not adequately serve the needs of PWD. But there are also specific ways in which AI systems, in particular those based on machine learning, could discriminate against PWD.

In many cases, ML systems may perform badly on PWD because they have not been exposed to a sufficient variety of people with those disabilities in the training set (Whittaker et al 2019). For instance, various biometric identification systems are trained on examples of physiological features of people without disabilities, and therefore fail to recognise people with disabilities. Biometrics based on the ways in which people without disabilities interact with their devices, such as typing and keystroke dynamics may not work as well or at all for people with conditions such as oligodactyly. A stark example is provided by automobile vision systems which cannot recognize wheelchair users due to the lack of examples in training data sets (or lack of examples identified by human labelers), with potentially lethal consequences (Nakamura 2019, Kraemer & Benton, 2015).

Features which are negatively correlated with a negative outcome (e.g. defaulting on a loan) in the general population may bear no such correlation among people with specific disabilities. For instance, some financial lenders report using machine learning models which have identified a positive correlation between correct capitalisation of words in loan applications and creditworthiness (Berg et al 2020). People with dyslexia might be unfairly downgraded as a result of such a feature in the model which may bear no relation to their actual ability to repay the loan.[90]

In other cases, AI systems may perform badly on PWD because historically, PWD have been discriminated against within the social contexts from which the training data are drawn, and such discrimination is reflected in the labels of PWD in the training data. Just because the training data does not include disability status, the algorithm may still pick up on proxies for disabilities.[91] For example, algorithmic systems used in the hiring process by many employers are fitted to historic data from the relevant profession, where examples of 'successful' candidates may be the result of structural discrimination such as exclusionary hiring practices or work environments (Madnani 2017).

---

[88] Trewin et al 2019, Guo et al 2019
[89] From the "*Call for Papers - Special Issue of ACM Transactions on Accessible Computing (TACCESS) on AI Fairness and People with Disabilities*" (https://dl.acm.org/journal/taccess/call-for-papers , archived at https://perma.cc/8259-E2C3)
[90] Whether it is fair to use such features for credit risk at all - even for people without disabilities - is a further question; see e.g. (Wachter & Mittelstadt 2020).

In some cases where AI systems are used to evaluate people, requesting accommodations such as more time may be a reasonable solution.[92] But for others, more time might not be sufficient. For instance in the context of AI facial analysis technology used in hiring, the candidate may have a disability which prevents them from presenting facial actions which resemble traditionally successful candidates.[93] In such cases, more time will not necessarily help.

The ethical concerns mentioned above can be described in terms of 'allocative' justice, in the sense that they primarily concern the allocation of goods, benefits, or services (e.g. credit, job offers). However, ML can also perpetuate representational or recognition-based harms which have to do with the way certain groups or identities are represented in culture (in this case, computational artefacts).[94] For instance, language models used in natural language processing and generation can reproduce stigma. In the context of disability, this means that language models intended to classify the sentiment or 'toxicity' of text could perpetuate harmful stereotypes about people with disabilities (Hutchinson 2020).

### 3.2. **AI fairness methods to mitigate or 'fix' fairness**

Various methods have been proposed to modify machine learning models so as to avoid unfair treatment based on protected characteristics. Simply removing protected characteristics from the data ('fairness through awareness') is typically ineffective because protected characteristics are correlated with other features used by AI systems. Instead, 'fairness through awareness' involves collecting protected characteristics and measuring performance on each group. There are two different classes of approach to fairness through awareness, namely: disparate learning processes and disparate treatment approaches. Since the distinction between the two has important consequences for equality and data protection law, as we will explain in section 4, we will briefly outline these two approaches here.

---

[91] Existing automated hiring systems which have made some efforts towards AI fairness often don't even consider PWD: "*examples of group definitions are binary gender, ethnicity (using definitions informed specifically by US demographics) and age interval. Social class or disabilities are not found in either the examples nor in the available validation tests*." (Sánchez-Monedero, Dencik, & Edwards 2020)

[92] See e.g. HireVue's approach to ADA compliance (https://hirevuesupport.zendesk.com/hc/en-us/articles/360028139512-ADA-Compliance-Steps, archived at: https://perma.cc/L6GS-QNQG).

[93] The use of such systems has faced staunch criticism beyond specific issues relating to PWD. HireVue claimed that *"facial actions can account for 29% of a person's employability score"* (https://www.washingtonpost.com/technology/2019/10/22/ai-hiring-face-scanning-algorithm-increasingly-decides-whether-you-deserve-job/, archived at https://perma.cc/UN4P-ELBH). Subsequently, in response to an audit by O'Neil Risk Management and Algorithmic Auditing, HireVue discontinued their use of facial action analysis (https://www.wired.com/story/job-screening-service-halts-facial-analysis-applicants/).

[94] See Solon Barocas and others (2017) 'The Problem With Bias: Allocative Versus Representational Harms in Machine Learning', Paper presented at the 9th Annual SIGCIS Conference, October 29 2017. http://meetings.sigcis.org/uploads/6/3/6/8/6368912/program.pdf (archived at: https://perma.cc/9X3R-A432) ; Kate Crawford, 'The Trouble with Bias' (NIPS 2017 Keynote) Available at https://www.youtube.com/watch?v=fMym_BKWQzk .

Disparate learning process approaches involve using data about protected characteristics (in this case, disability) during the *development* of the AI system, but not during its *deployment*. The simplest involve removing any features in the training set that correlate with a protected characteristic, until the resulting model complies with the chosen fairness metric. However, these typically reduce the accuracy of the model and thus its real-world performance. More sophisticated approaches involve using the protected characteristics to create and impose a constraint on the learning process without having to remove correlated features altogether.[95] Similarly, 'fair representation learning' involves pre-processing training through an encoder which removes information about protected characteristics (including proxies) such that the input is still useful for prediction but information about the protected characteristic is suppressed. In each case, the resulting model is one which does not use protected characteristics as an input for new classifications and predictions, even though they were used in the model development process. This means that while protected characteristics need to be collected for the development of the ML system (typically, for (a subset of) the individuals in the train / test data), but not for real users during deployment. The following sections refer to these approaches collectively as DLPs.

By contrast, treatment disparity (TD) approaches involve using protected characteristics not only in the development / training of ML models, but also directly as a feature in the model and therefore also as an input to the model during deployment. 'Treatment disparity' therefore reflects that these models treat individuals differently depending on their protected characteristics (the term 'within-group scoring' is also used).[96] Treatment disparity approaches have several advantages; they are able to strike a better trade off between accuracy and group parity metrics; they preserve the rankings within protected groups, and they do not introduce within-group disparities (Lipton, McAuley, & Chouldechova 2018).

These algorithmic fairness methods are themselves contentious for a variety of reasons. In general, they have been criticized for using a notion of 'bias' or 'fairness' which is divorced from the conditions of societal injustice which produces 'biased' training data, which may, once deployed, only serve to re-entrench those pre-existing conditions and the power of dominant groups in society (e.g. Greene et al 2019; Hoffman 2019; Binns 2018). They also borrow loosely from U.S. models of anti-discrimination (which are in themselves problematic (Hoffman 2019)) and apply them in a homogenizing way across different axes of discrimination (treating e.g. race, gender, disability, etc, in the same way). In the context of disability specifically, they may fail because they typically require the simplifying assumption that the relevant axes of discrimination are neatly categorisable. While this assumption is wrong for all protected categories, which are all internally diverse, it is especially so for people with disabilities. Disability as a category is especially internally diverse; as Trewin argues: 'compared to gender, race or age, it is not easy to address biased outcomes by gathering a balanced set of training data, because there are so many forms and degrees of disability' (Trewin 2019). Finally, these problems are all compounded in the case of PWD, since the legal concepts involved in disability law are themselves often distinct from other areas of discrimination law.

These problems notwithstanding, in at least some cases – where there is sufficient data on PWD that the model can be modified, where privacy is adequately protected, and where other structural measures are developed and

---

[95] See e.g. Zafar et al 2017; Lipton, McAuley, & Chouldechova 2018; Pedreschi et al 2008
[96] See section 5.2 of Raghavan et al 2019

maintained – algorithmic fairness methods (whether DLP or TD-based) could have a meaningful role to play in mitigating harms for PWD. Other approaches to AI fairness might also be explored. For instance, systems could be designed to provide "*manual overrides for outlier individuals where the model is unreliable*", or individuals from different disability groups could be routed to alternative models which have been designed to perform better for them, e.g. "*a specialized speech recognition model tuned to the characteristics of people with slurred speech, or people who stutter*" (Trewin et al 2019). Trewin et al caution that this would require careful thinking, and presents a particular 'ethical minefield' if done on the basis of AI-*inferred* disability. While different in spirit, these approaches are similar to treatment disparity approaches in that they involve explicitly using an individual's disability status in order to treat them differently, so raise similar ethical questions.

However, many researchers and practitioners are reluctant to recommend or use treatment disparity approaches for fear that they may constitute unlawful disparate treatment (in the US) or direct discrimination (in the EU).[97] Even if they turn out to be effective means of avoiding indirect discrimination, they appear to do so in a way that might constitute direct discrimination. They also involve collecting protected characteristics during deployment, which individuals may (often understandably and justifiably) object to.

### 3.3 Summarising the current state of AI Fairness for PWD

As explained above, the current state of AI fairness research encompasses a range of fairness or justice-related concerns regarding the harms of AI systems, and involves a wide range of disciplinary approaches and divergent political outlooks. While methods for mitigating or 'fixing' unfair AI systems have been proposed, there is substantial debate as to whether such methods might be required, permitted or prohibited by law. However, in so far as these discussions have considered discrimination law, they have looked at protected characteristics in general. Typically, they have not focused on disability in particular, which, as discussed above, is treated differently to other protected characteristics in many legal jurisdictions and involves a range of distinctive risks and practical considerations.

We believe that researchers and practitioners working in this space will make progress by considering discrimination law more holistically in conjunction with other areas of law, as a productive way of understanding of opportunities and barriers. Stakeholders involved in deploying (or contesting) AI systems are not the first to grapple with these issues; after all the law is not silent on the ethical minefield surrounding differential treatment of people. In the context of this paper, that conjunction is disability and data protection law. The following section discusses how equality and data protection law apply to these questions in the context of AI and PWD.

### 4. WHAT KINDS OF AI FAIRNESS APPROACHES DO EQUALITY AND DATA PROTECTION LAW REQUIRE OR ALLOW?

We now turn to consider the implications and effects of both data protection law and equality law on AI fairness and PWD. In particular, we address concerns around identifying the positive obligations that may apply to developers and/or deployers of AI systems with regard to PWD. At the same time, we consider the extent to which they may allow, require, or perhaps even preclude the kinds of algorithmic fairness approaches we outlined in the previous section. While the legal details may not be a primary concern for AI practitioners, they are important because they constrain, shape and support any agenda for 'AI fairness'.

---

[97] For examples of those concerned that treatment disparity would violate disparate treatment, see e.g. Barocas & Selbst 2016; Grimmelmann & Westriech 2016; Nachbar 2020. For arguments to the contrary, see e.g. Kim 2016; Hellman 2020. For discussion, see Lipton, McAuley & Chouldechova 2018; Xiang & Raji 2018.

In order to illustrate these implications, we use a running example of a hypothetical employer who uses an AI service to analyse and score video interviews for recruitment purposes.[98] To illustrate the implications for a specific class of disability, we will use the example of applicants with speech impairments. While such systems have a wider set of ethical and legal problems, we use them here as an example to illustrate the application of equality and data protection law with respect to PWD more specifically.[99] It is plausible that such systems would mistake the effects of a speech impairment with patterns of speech that are associated with worse candidates in the training data. Candidates with speech disorders might therefore be 'unfavorably treated' by AI-based recruitment video analysis.

### 4.1. Equality law, AI systems and PWD

In principle, Equality Law requires that a system does not discriminate, provided it is applied in one of the broad domains covered by the Equality Act (2010) (with Services, Public Functions, Education and Work being the main areas of concern for AI). Delegating decision making to a computer does not relieve any general obligations of the Act: an organisation relying on an AI system (either fully, or to support humans in decision making) still must ensure that it is not taking discriminatory decisions.

On a simple level, this means the **duty not to discriminate on the basis of all protected characteristics *when taking decisions***: a system must not directly discriminate or indirectly discriminate. This framing of discrimination, as a property of decision-making processes, is typically what is adopted in existing 'AI fairness' literature. In other words, it must not take a decision that is less favourable because of disability, and in respect of factors related to disability, it must not treat someone less favourably due to those factors, unless that can be objectively justified. In the AI-based recruitment video analysis example, this could be established by comparing the outputs of analysis for applicants with and without speech impairments. However, this duty goes beyond decision making and applies to any PCP that may put any group of PWD at a disadvantage: for instance, some PWD will be especially concerned with the **confidentiality and management of disability status data, so the usage of any system that did not assuage these concerns would be indirectly discriminatory**, unless that could be objectively justified.[100] So, when considering the use of or implementation of an AI system for people with disabilities, there has to be a **wider and holistic consideration of potential implications**, if an organisation is to ensure it meets its obligations: it is not enough to purely focus on the outcomes of the decision making process itself.

Similarly, **where the service is not itself based on AI, then there is a duty to make available any reasonable alternatives** for individuals who, due to their disability, would be placed at a "*substantial disadvantage*" due to a PCP (or at a detriment arising from due to "*something*" related to their disability): however, there is seemingly no duty to alter the service itself if it is inherently an AI system (compared to an organization that is simply applying an AI system in its day to day operations, e.g. to help sell flights that it operates). Furthermore, unlike with other protected characteristics, it is **perfectly lawful (and we expect obligatory) for an organisation to deliberately treat PWD more favourably**, which is a route around the difficulties that direct discrimination poses for other domains of AI fairness. Treating PWD differently and more favourably is often an important feature of disability discrimination law, and this is no different when considering the application of AI to (and sometimes for) PWD. In the context of the AI-based recruitment video analysis, this might mean weighting or discounting the score outputted by the system for individuals identified as having a speech impediment (the data protection implications of such identification are explored below).

---

[98] For an overview and real examples of such services which include automated video analysis, and their equality and data protection implications, see (Sánchez-Monedero, Dencik, & Edwards 2020).
[99] Ibid 93
[100] Or there was another exemption under the Equality Act (2010), or alternatively, it concerned a matter that wasn't within its ambit..

There are two other important considerations with respect to AI fairness. First, there is an **active evaluative duty** where an organization is a state entity subject to the Public Sector Equality Duty, which is active and investigatory.[101] In practice, we observe that any organisation in the private sector would almost inevitably not comply in some way without also actively investigating compliance of an AI system (with this being especially likely for PWD, given the wider obligations that apply and the diversity within this group); thus there is an implicit duty on these organisations to engage in an evaluation of any AI system they wish to deploy. In our example, the recruiter may therefore need to evaluate AI-based hiring systems prior to procurement and use.

Second, the wider importance of disability equality, and the persistent barriers that face many PWD, including the relatively limited quality of life (due to existing social factors) is a **considerable weight that can be considered in any relevant proportionality exercise**, including when taking a relevant decision under data protection law. This means that any balancing exercise should take into account the bigger picture of the social obligations and expectations of Equality Law, and the need to ensure that it works more effectively on the ground. It also means considering the risk of potential discrimination, not just discrimination that can be shown to take place. Indeed, as we will explain, data protection law could in some cases be used to circumvent the limitations in enforcement that have hobbled the effective implementation of the Equality Act (2010) on the ground.

### 4.2. Data protection law, AI systems and PWD

Broadly, efforts to promote fairness for PWD in the context of AI depend in various ways on collecting, processing and analysing data about PWD. One of the challenges here is concerns around privacy and use of data. As Morris argues, some PWD may withhold their disability status due to (often legitimate) privacy fears, which can 'further amplify the inclusivity problem of AI systems' (Morris 2019). As Givens and Morris note, part of the solution may involve the creation and curation of 'inclusive datasets', which raises questions about the ethics of collecting and centralising disability data, and protecting the identities of PWDs in such datasets (Givens & Morris 2020). Researchers and organisations may need to collect and use data about people with disabilities in order to pursue AI fairness ends, for instance to test possible differential performance of deployed AI systems for PWD.

---

[101] See Bridges, R (On Application of) v The Chief Constable of South Wales Police [2019] EWHC 2341 (Admin), where the PSED's applicability to AI systems was highlighted. More generally, it's worth observing that in respect of the states international obligations, the UN CRPD requires statistics and data collection under Article 31 to monitor the wider implications of policies and practices on PWD, including those by non-state entities. Arguably this means that states should be actively considering all AI systems deployed in wider society, not just those they operate themselves, or rely upon to support their own decision making.

Ideally the law should support such efforts to collect disability data where they are ethical, but it may also inadvertently present barriers. Various researchers working on AI fairness for PWD have raised general concerns about how privacy and data protection laws may inhibit the ability of those deploying AI to measure and mitigate its potentially unfair treatment of PWD. While some of these concerns are in response to U.S. laws,[102] others are particularly concerned with EU data protection. Trewin notes that the GDPR may mean that as "*organizations move to limit the information they store and the ways it can be used ... AI systems may often not have explicit information about disability that can be used to apply established fairness tests*" (Trewin 2018).

To respond to these concerns, this section outlines how data protection law applies to the collection of disability data to enable AI fairness. In all of these cases, data protection law already provides some safeguards which limit the ways in which such data can be used, but also enable certain uses within those constraints. These safeguards vary slightly depending on whether data revealing an individual's disability is used as part of a solely automated decision-making process or not. We begin by addressing cases of decision-making using disability data which do not involve using it as an input to a solely automated decision.

4.2.1. *Using disability data for AI fairness*

As explained in section 2.2, data revealing someone's disability status is 'special category data' under the GDPR. This means that an organisation wishing to process it must have both an appropriate lawful basis, as well as meet one of the special conditions under Article 9. Various Article 9 conditions may be appropriate depending on the circumstances, however the following examples illustrate some plausible options for this context.

Perhaps the simplest condition under which special category data can be processed is with the data subject's explicit consent. However, as mentioned above, this consent would need to be 'freely given'; this would be undermined if consent is made a precondition for accessing a service, or if the data controller is in a position of power over the data subject (e.g. if they are a public body responsible for delivering an essential service to the data subject, or the data subject's employer). This means that the cases where consent could be relied on are quite limited. Alternative Article 9 conditions may therefore be preferable. These include where the processing is carried out as part of the controller's obligations under employment, social security and social protection law; by a not-for-profit body where the data concern members or persons they have regular contact with; where there is a substantial public interest; for health or social care by relevant professionals; or for research purposes.

For data controllers who are pursuing AI fairness for PWD, whether as researchers, or as employers, public bodies, or other organisations actually deploying AI systems, any one of these conditions would be sufficient to comply with Article 9 and lawfully process special category data (in this case, disability status). Specifically, such Article 9 grounds could enable the pursuit of AI fairness, where data containing disability labels need to be processed to inform fairness analysis. In our AI hiring example, the provider of the video analysis software could compare model performance between people with and without speech impediments and adjust the model accordingly; recording disability status in order to make such comparisons in accordance with Article 9.

---

[102] For instance, in the U.S., the Americans With Disabilities Act governs what questions employers are allowed to ask about a candidate's disability (Raghavan 2019), while the Health Insurance Portability and Accountability Act (HIPAA), and the Genetic Information Privacy Act regulate the sharing of health information (Givens & Morris 2020).

### 4.2.2. *Solely automated decisions*

However, the situation is more complicated for controllers taking any approach which involves using disability directly as an input to decisions made by the system with regard to individuals themselves during live deployment (as with, for example, the treatment disparity approaches, or approaches which involve routing PWD to specialised alternative models). As explained in section 2 above, processing of personal data is constrained where it is used for solely automated decisions with legal or similarly significant effect, and constrained even further where those decisions are based on special category data. To return to our example, an employer who used data about a candidate's speech impediment as an input to an automated interview screening process (e.g. in order to weight or discount the score used to automatically determine whether they are invited to interview), they would not only need an Article 9 ground, but specifically one of the subset of Article 9 grounds specified by Article 22(4).

So for cases where a system involves solely automated decision-making, and an individual's disability data is used as an input to the decision, a data controller can only rely on explicit consent, or one of the substantial public interest grounds.[103] However, these conditions are not always going to be appropriate or easy to fulfill. Consent will often not be appropriate, for instance where there is a power imbalance between the data subject and the data controller. This would be the case in employment contexts.

This leaves only the substantial public interest condition. Recall that what legitimately counts as 'substantial public interest' must be defined in member state law. Under the UK Data Protection Act Schedule 1, there are 23 possible substantial public interest conditions. While some of these may appear at first glance to be appropriate for the purposes of AI fairness for PWD, on closer inspection they may turn out to be inapplicable in context. For instance, Schedule (1)8 on 'equality of opportunity or treatment' might seem appropriate, but could not be relied on here because clause 3 of that condition clarifies that the condition is not met if the processing is 'carried out for the purposes of measures or decisions with respect to a particular data subject'. This would appear to rule out DT approaches which involve using protected characteristics as an input to decisions about individuals (unlike DLP approaches, which don't use the protected characteristic as an input directly in the decision).

However, some of the conditions do appear to enable DT approaches for at least some kinds of AI fairness efforts. Where there is no power imbalance between the data subject and controller, consent remains a possible lawful basis, as well as well as being one in the subset of Article 9 conditions that would enable disability to be used as a direct input to a solely automated decision according to Article 22(4). Schedule 1 para 16 allows special category data to be used, even to make a decision about an individual, if the processing is carried out by a not-for-profit body which provides support to individuals with a particular disability or medical condition.

---

[103] Article 22(4): "*Decisions referred to in paragraph 2 shall not be based on special categories of personal data referred to in Article 9(1), unless point (a) [consent] or (g) [substantial public interest] of Article 9(2) apply*"

This analysis suggests that while data protection does strongly constrain the contexts in which disability data can lawfully be processed, it heavily depends on whether such data is used to merely test and constrain AI systems according to fairness constraints, or as an input to solely automated decision making under Article 22. If the former, there are multiple avenues through which different organisations in various capacities - including researchers, disability support groups, employers, and others - may legally collect disability data for some AI fairness purposes. The appropriate lawful basis and processing conditions will clearly depend heavily on the context, but there is flexibility to support a variety of approaches from different stakeholders, and this may even encourage fruitful collaborations between academic researchers, disabled people's organisations, employers and other organisations. As a result, data protection should not be a barrier to such efforts, at least in the UK and other jurisdictions with similar laws.

However, where data controllers wish to use disability status directly as an input to an AI system to ensure fair treatment of PWD, the options are tightly constrained; they must either obtain explicit consent of the individual (which may be inappropriate where there is a power imbalance), or fulfill one of the public interest grounds (which are quite narrow). If those options are not viable, and the only way to make an AI system 'fair' is to directly use disability as an input, then the organisation may find itself in a bind. On the one hand, by not using disability as an input to the solely automated decision, it may end up indirectly discriminating against PWD contrary to equality law; but on the other hand, if it does use disability as an input, and neither explicit consent nor any of the substantial public interest conditions are applicable, then it could be unlawful under data protection. In such cases, the combination of data protection and equality law could make it impossible to use AI systems lawfully.

In some ways, this may be a desirable outcome; despite being well-intentioned, algorithmic fairness methods have important shortcomings and could provide a false sense of technically 'fixing' more fundamental problems. Their limitations with respect to disability in the context of data protection law may add force to existing critiques of these approaches as relying on a "*limited and essentialist reading of fluid and socially constructed categories of identity*" (West, Whittaker & Crawford 2019). If the combination of these areas of law does not support their use for legal or similarly significant and solely automated decision-making, and this prevents the deployment of harmful AI systems with insufficient algorithmic fairness fixes, this may present a welcome opportunity to re-consider such approaches. At the same time, in some contexts PWD may appreciate the opportunities that carefully 'fairness-adjusted' AI systems might afford them; it would be a shame if the opportunities to explore algorithmic fairness approaches in such contexts were hindered by laws which were not drafted with these approaches in mind.

4.2.3. *GDPR obligations and individual rights promoting AI fairness*

As well as setting out some conditions under which organisations can lawfully process disability data and make solely automated decisions about PWD, data protection may also in some cases, arguably, even *require* organisations using AI systems to undertake various AI fairness measures. For instance, recital 71 discusses rights and protections in relation to automated decision making and profiling, including requiring data controllers to implement technical and organisational measures to minimise inaccuracies and errors, and to prevent discriminatory effects on the basis of special category data (which would include disability).[104] The GDPR even has an overarching principle of fairness, which is broad and not precisely defined, but has been interpreted to include avoiding unjustified adverse effects on individuals.[105] More broadly, in so far as data protection aims to protect the fundamental rights and freedoms of data subjects, including non-discrimination, and in so far as disability discrimination can be perpetuated by AI, data protection should support rather than hinder efforts to reduce discrimination against PWD.

---

[104] While recitals are not operative provisions of the GDPR, they clarify and expand how the operative provisions are to be interpreted.

4.3. **Summarising the qualities of Equality Law and Data Protection Law that relate to AI Fairness for PWD**

The foregoing account has illustrated why equality law should be considered alongside data protection law for the purposes of AI fairness, as well as the particular circumstances where these concerns intersect. It is apparent that challenges that might deceptively appear to be issues of equality law might often be more strategically addressed by accessibility advocates by focusing on data protection legislation. **Table 1** below provides a summary of the perspectives on key issues provided by both legal approaches, and the key provisions within them that apply to topics of concern. In the final column of **Table 1**, we highlight the potential tactical implications for accessibility advocates, which we develop further in Section 5 as an agenda for accessibility research and advocacy.

| Obligation or Ability | Equality Law Perspective | Data Protection Law Perspective | Key distinctions and implications |
|---|---|---|---|
| Security and protection of disability related data | Failing to do this would likely amount to (i) indirect discrimination, (ii) a failure to make reasonable adjustments and (iii) discrimination arising from a disability. | Processing of 'special category' data (which includes disability) requires a lawful basis and an Article 9 condition; such data must also be secured and data breaches must be reported. | The duties to secure and protect disability related data in data protection law are well defined and more specific than in equality law. This means that a focus primarily on data protection law should be adopted when considering how to protect against detrimental data processing. |
| Taking a less favourable (automated) decision because of disability status. | If this applies to all people in a disability group, then it is direct discrimination, otherwise in most circumstances, it would be indirect discrimination and discrimination arising from a disability. | There are limited conditions under which solely automated decisions with legal or similarly significant effect can be made; these are even stricter when disability data is used as an input. This is irrespective of whether the decision is adverse. | Equality law is the primary means of addressing discrimination, whether automated or human, while data protection law puts extra safeguards around solely automated decisions, whether discriminatory or not. Thus, both areas of law should be considered both separately and cumulatively when attempting to protect PWD from less favorable treatment. |

---

[105] See 'Principle (a): Lawfulness, fairness and transparency', Information Commissioner's Office, https://ico.org.uk/for-organisations/guide-to-data-protection/guide-to-the-general-data-protection-regulation-gdpr/principles/lawfulness-fairness-and-transparency/

| Taking a more favourable approach to some people with disabilities than those without. | This is always permitted for disability, unlike with other protected characteristics. In some circumstances, it can be required to do so as a reasonable adjustment. | DP law does not generally treat disability differently to other protected characteristics, although there is one condition specifically allowing non-profit bodies to process disability data in this way. | The differential treatment of disability in equality law (compared to other 'symmetrically' protected characteristics) is only partly reflected in data protection law. This means there is a risk of information practices based on DP law overlooking the duty to make reasonable adjustments, which should be considered carefully by accessibility advocates. |
|---|---|---|---|
| Duty to provide alternatives to AI approaches if a PWD subject to a substantial disadvantage. | This can apply, if it's a reasonable adjustment to do so (i.e. there is a strong enough case). However, this does not apply if it would involve altering the fundamental character of a service. | DP law provides the right to contest solely automated decisions (e.g. those made by AI systems) with significant effects, and to request a human intervention; this applies regardless of whether the automated process can be shown to disadvantage the individual. | Alternative decision processes must be provided under both equality and DP law in certain circumstances. This means that both approaches offer different opportunities for protecting PWD, both of which should be considered by disability advocates. |
| Duty to investigate and actively avoid possibly discriminatory practices. | This applies directly to public authorities. In some cases there can be civil and criminal implications for professionals who promote discrimination. | Data protection impact assessments are required for risky processing (such as AI and large scale processing of disability data) and cover risks to rights and freedoms, including disability discrimination and outline measures to mitigate that risk. | Some forms of ex ante investigation and mitigation of AI (un)fairness are required under both equality and DP law, but in the former case this applies only to public authorities, and in the latter case only to 'high risk' processing. When looking at encouraging organisations to act proactively (rather than reactively), the avenues offered by DP law and equality law should be separately considered. |
| Enforcement of discrimination law. | In most cases, this is by individual action and proving that one was discriminated against. | Data protection aims to prevent discrimination indirectly through requiring organisations to consider fundamental rights when designing systems for processing personal data. | While equality law is the primary vehicle for discrimination cases, pursued through litigation, data protection plays a supporting role by requiring such rights to be embedded by design. As such, there are many cases where disability rights might be best enforced by considering a DP law angle, rather than focusing on equality law. |

*Table 1 – An overview of differences between data protection law and equality law.*

5. **USING DATA PROTECTION AND EQUALITY LAW TO PROMOTE AI FAIRNESS FOR PWD IN THE DESIGN AND IMPLEMENTATION OF (ASSISTIVE) TECHNOLOGY.**

This section considers some of the ways in which data protection and equality law (including disability human rights law) could play a role in the development of AI fairness for PWD in the near term. Equality and data protection law could be used in different ways according to differing political approaches to addressing the potential harms of AI for PWD. For instance, for those who believe in reforming AI systems to be less harmful or even positive for PWD, these areas of law can provide relevant exemptions, conditions, and incentives to promote such reforms. At the same time, for those who are skeptical of 'AI fairness' approaches, and pursue an abolitionist approach towards a given AI system (e.g. Bennett & Keyes 2019; Benjamin 2019), data protection and equality law may still provide some tools through which objectively inappropriate AI systems might be resisted and challenged. As with all areas of law, they contain *both* problematic elements of the political status quo, *and* emancipatory potential; as such, they should neither be naively embraced as a panacea nor dismissed as inherently ineffective. In the spirit of a 'critical jurisprudence of disability', we acknowledge that the law and legal institutions may unfortunately all too often still contain 'overt and covert sources of oppression' against PWD; whilst also recognizing the 'potential positive role of law … in the struggle for the emancipation of disabled people' (Hosking 2008).

We argue that considering these two domains of law together could help advance the interests, rights and opportunities of PWD in respect of AI. In some cases, this might involve using these laws to challenge the deployment of AI systems which harm PWD; but they may also play a role in encouraging the design and evaluation of assistive technology for PWD, and in helping advance the universal design of more generalist AI systems to be more inclusive of PWD. Particularly notable are the opportunities that these provide to both disability rights organisations and assistive technology researchers alike to advance the rights of PWD going forwards. We outline three pertinent areas, namely:

- Challenging the deployment of harmful automated decision making systems
- Setting an agenda for data protection law-compliant automated assistive technology to help prevent discrimination against people with disabilities
- Advancing the role of disability rights organizations in the governance of AI

The overall result is to take what we have learned in the previous sections and explain how this can be put into practice by accessibility advocates, researchers and practitioners on the ground. We do this in two parts: the first is an overview of how harmful automated decision making systems can be directly challenged (in 5.1), before moving to propose a wider agenda for assistive technologies researchers.(in 5.2).

5.1. **Challenging harmful automated decision-making systems:**

The rights provided under data protection law point to a variety of ways that PWD could challenge harmful AI systems. In fact, many of these rights correspond directly to those which have been called for in recent work on AI fairness for PWD. For instance, Trewin et al argue that where AI systems are used to make decisions about them, PWD should have the right to: "*inspect and correct the data used to make decisions about them*"; "*query and challenge*

*AI decisions, receiving some form of explanation of the factors that most impacted the decision*"; and have the "*opportunity to dispute a decision, and provide a manual override for outlier individuals where the model is unreliabl*e".

Each of these rights are already enshrined in some form in the GDPR. The right to rectification (Article 16) gives data subjects the right to have inaccurate data corrected, or to have 'incomplete' data completed by means of a supplementary statement. This means that where an automated system is used to make an inference or evaluation about a person, the individual can ask for the input data (which they may request a copy of under Article 13), to be corrected where it is inaccurate as to a matter of fact. While this doesn't necessarily extend to inferences themselves (because they may be more akin to opinions rather than matters of fact), rectification would give the individual the right to put forward information that would contextualise those inferences, or indeed provide evidence that contradicts said inferences. Being able to put AI inferences in context in this way could be crucial for PWD, where systems might systematically underestimate their true capabilities due to a disability-insensitive AI model.

Where decisions about individuals are made using AI in a solely automated way, with legal or similarly significant effects, the individual also has under Article 13(2)f the right to meaningful information about the logic involved, and under Article 22(3) rights to contest the decision, to request human intervention, and to put across their own point of view. These latter rights under 22(3) mean that organisations must put in place the capacity to manually override AI systems, including having human decision-makers on hand to reconsider the decision and to respond to cases where it is contested.

While important, these rights are somewhat limited in so far as they rely on individuals having the resources and capacity to exercise them (albeit this might often be a lot easier to do than to bring a claim for discrimination in the County Court using Equality Law). As such, any benefits they provide may go to those PWD who are already relatively advantaged in various ways, and risk leaving out those who most need redress from AI-based harm. However, despite being oriented towards individual rights, data protection law nevertheless has various mechanisms through which AI systems can be challenged and shaped to protect all PWD at a more systemic level by disability rights groups, trade unions, or other civil society actors although this is often under-considered in respect of technological innovations for people with disabilities (Alper, 2017). What's more, the AT community is well positioned to thread the needle between data protection law and disability rights, developing a well-evidenced body of principles, which can ease any action taken by these organisations. Similarly, the AT community's long experience of developing heuristics and principles could lead to a voluntary code and effective standards[106] (like WCAG) which influences practice on the ground, thus minimising the need for legal recourse.

For cases that are more contentious, one possibility for advancing the treatment of PWD is for civil society organisations - ideally with the support of evidence and best practice developed by AT researchers - to challenge the lawfulness of data processing through the courts. Recent years have seen several organisations challenge algorithmic systems using Article 80 of the GDPR which allows non-profit organisations to take cases on behalf of data subjects. One example is the SyRi algorithm used by the Dutch government to detect possible welfare fraud, which was successfully challenged by a coalition of digital rights groups and trade unions as unlawful under several grounds in the GDPR and human rights law.[107] While data protection law typically doesn't ban particular technologies outright, it provides a framework within which certain uses of certain technologies may be found unlawful because they do not comply with the procedural safeguards, or where they involve processing of personal data which disproportionately interferes with rights and freedoms of data subjects. As with the SyRi case, data protection can often be an initial step in a broader challenge to the lawfulness of an entire system. Furthermore, an AI system that facilitates discrimination could lead to civil and criminal penalities for the individuals involved under Equality Law (including the designers of such a system), providing a further incentive for compliance.[108]

---

[106] Standards are well known to be of critical importance when increasing the accessibility of technology (see for example, the accounts of Brewer, 2017, and Blanck, 2014).

In addition, data protection law does provide some opportunities through which collectives can exert influence over the design of AI systems, and potentially even over whether they are deployed at all. A case in point is that data controllers are required (under Article 35) to conduct mandatory data protection impact assessments prior to processing personal data in new high risk ways (such as AI). As part of this, in some cases controllers must consult data subjects or their representatives to understand the risks and possible measures to mitigate them. Unhelpfully, this requirement is vague (only applicable 'where appropriate', which has yet to be clarified through case law), but is nonetheless another entry point for challenging AI systems before they are deployed.

### 5.2. An agenda for data protection law-compliant automated assistive technology to help prevent discrimination for people with disabilities

Below, we set out specific examples of what can be accomplished by applying data protection law to further the design of assistive technology, and to advance the wider interests of PWD.

#### 5.2.1. *Challenging generally discriminatory practices and (automated) systems*

As we have observed earlier in this article, disability discrimination remains widespread, with serious social consequences for many PWD. It is difficult to understate the importance of this concern, given how widespread disability discrimination remains in today's society: any tools that can help evidence discriminatory practice (and serve as the basis of challenge) could greatly increase the life opportunities of PWD. At present, there is relatively little data on whether or not most practices present a substantial disadvantage (or, less favourable treatment), making it harder to challenge them. By way of an example, existing practice in administrative tribunals that assess disability has been described as being based on "*anecdote and supposition*"[109] despite them dealing with 100,000's of disability benefits claims in the UK in hearings behind closed doors. The same point might be said about most other widespread public sector practice, let alone those peculiar to smaller organisations. Thus, if practices and automated systems were able to be identified and evaluated more effectively, then this would help reduce the level of discrimination against people with disabilities. These evaluations could arise in respect of two different scenarios:

---

[107] NJCM cs/ De Staat der Nederlanden (NJCM vs the Netherlands): https://uitspraken.rechtspraak.nl/inziendocument?id=ECLI:NL:RBDHA:2020:1878 (English translation). Archived at: https://perma.cc/G2N5-LADV
[108] In the UK, this would be via Part 8 of the Equality Act (2010), as set out in Section 2.1. of this article.
[109] By no less than the person ultimately responsible for that system: https://www.judiciary.uk/wp-content/uploads/2019/09/2019_09_19_SPT_SLS_Preston_Sept19_FINAL.pdf (archived at: https://perma.cc/B3Z3-VT2S )

1. By organisations themselves seeking to find out if their own practices or automated systems may discriminate against people with disabilities (or a group thereof). After all, many organisations would prefer to avoid discriminating against PWD, and much discrimination is unintentional, rather than deliberate.

2. By assistive technology researchers and other advocates (including disabled persons organisations) seeking to conduct a more general investigation of potential discrimination, or to establish the least discriminatory (or risky) approach that can be adopted.

We observe that these evaluations can take place at a population level, without needing to have identifiable individuals in a dataset sent off for analysis. So, in addition to facilitating organisations' internal compliance efforts, data protection law should also not necessarily present a barrier to disability advocates and activists seeking to undertake population level analysis of potentially discriminatory practices. Furthermore, by putting place safeguards to protect PWD who do choose to share sensitive personal data, data protection law may allow for disability advocates and activists to have new opportunities to challenge AI systems, and thus enable them to bring substantial (and otherwise invisible) injustices to account.

At the same time; automated approaches for evaluating discrimination in general that are designed to fit within an already established set of legal boundaries are far more efficient than previous approaches, thus being practical from a resource perspective for (often very hard pressed) disability rights organisations and their supporters. Evaluating discriminatory practices has been an issue that the AT community has touched upon (Lazar 2010; Lazar 2016): in this vein, tools that assist PWD (and advocates) to evidence discrimination are themselves a form of assistive technology, and it's worth considering the opportunities presented by DP law as a means for using AI to *advance fairness* by creating AT that leverages these new possibilities, rather than just considering if AI i*s itself* 'fair'. After all Equality law provides extensive (on paper) protections for PWD which are often not put into practice on the ground, due to the burdens it poses for claimants. As such, we argue that supporting general evaluations, and providing tools for PWD (and disabled persons organisations) should be pursued more actively going forwards by leveraging data protection law, and creating a trusted space for these types of evaluations to take place.

5.2.2. *Enabling the trusted sharing of disability-specific data*

The sharing of disability-related data has a double-edged quality to it: whilst there are great potential benefits (e.g. the data could be used to foreground more inclusion for people with disabilities), there are also many potential disadvantages (such as invisible discrimination by insurance algorithms) (Whittaker et al, 2019). For many PWD, disclosing information about disability is often avoided, especially those with invisible impairments, often due to (an often justified) fear of unlawful discrimination, or a lack of discretion as to onward disclosure. Perhaps the most important contribution of data protection law is the trusted space it can create for the sharing of personal data. Provided that the GDPR and the protections it offers are implemented robustly by an organisation, and effectively enforced by regulators, then these protections for special category data could provide justified confidence. At the same time, automated decision making or data analysis has the potential advantage of not requiring a human to view the information, thus providing a process that some people with disabilities might be more willing to engage with.[110]

Notably, the GDPR's structured rules in respect of special category data (including disability status), make it more difficult for any employer or organisation to successfully claim that they cannot assess the potential harms of their AI systems because doing so would involve unreasonable or disproportionate collection of sensitive data, because the legislation provides an explicit legal route for managing this data. We observe that this is an important example of the GDPR giving effectively more force to the reasonable adjustment duty, by raising expectations as to what an organisation must do to begin with (and thus making it a lot more difficult to claim that doing so would be unduly burdensome, and in turn not legally reasonable).

Thus, the creation of trusted spaces for disclosure of disability information by way of data protection law – including for automated consideration – is an important new opportunity that AT researchers should consider. The forms that such trusted spaces could take are varied, but could include 'data trusts' and intermediaries set up for such purposes, separated from commercial exploitation, registered with and monitored by public authorities. Indeed such alternative forms of data governance are envisioned in newly proposed EU regulatory instruments.[111] Such spaces could be useful for the design process itself, and AT researchers and designers could play a role alongside other stakeholders in stewarding disability datasets to be used for testing within the AI development process. But they could also become part of standard, on-going, iterative anti-discrimination assessment in deployed operational systems. Indeed, these trusted spaces undergird the other opportunities that we raise, including in respect of 5.3 (our discussion of disabled persons organisations).

5.2.3. *Supporting people with disabilities to make disability-related claims regarding AI systems*

Whilst bringing individual claims relating to disability discrimination can be challenging, it is important to note that AI could help assist with evidencing both the effects of disability, as well as if a practice has an influence on an individual's disability. A difficult issue surrounding disability discrimination is the need to gather evidence; a strong claim for discrimination has to show individual disadvantage to a *specific claimant*. A similar point applies to many other decisions: for example, to claim a disability benefit, someone with a disability has to prove they are sufficiently disabled (against often arbitrary formal criteria), which is a concern that greatly impacts on a large number of disabled people (see Watson et al 2020 for a discussion of this). There are two aspects of data protection law which could help support such endeavors:

First, some of the individual transparency rights under the GDPR could be used to obtain evidence that might help with this concern. For instance, the right to access a copy of all personal data held by a data controller (Article 13) would include any data provided by the data subject directly, observed behavioural data, intermediate transformations of such data that serve as inputs to an AI system, and any further data derived from the AI analysis which is associated with the individual. This could reveal, for example, how the collection and manipulation of data before or during the AI processing may result in unfavourable treatment for that individual on the basis of their disability. For instance, employers who use AI services to analyse and score video interviews for recruitment purposes would be obliged to share with an applicant a copy of the interview itself and the resulting score. This could help an applicant who may have been scored unfavorably as a result of a speech impediment evidence their claim.[112]

---

[110] Again, such willingness would of course be dependent on the development of actually trustworthy and non-harmful AI systems.
[111] 'Regulation on data governance – Questions and Answers', subsection: 'How will data intermediaries ensure trust in data sharing?' https://ec.europa.eu/commission/presscorner/detail/en/QANDA_20_2103 (archived at: https://perma.cc/T85T-Z2LX)

Second, the obligation to provide meaningful information about the logic of automated decisions under the GDPR (Article 13 (2)f) is also important, because it could indirectly mean that the evidence gathered by a PWD is more likely to be persuasive to a court, tribunal, or other decision maker, who have difficulties in accepting such evidence (see e.g. Weinberg, 2020 for an example of this concern). Such explainability might also perhaps help to protect against misunderstandings of generic disability measurement systems that someone might repurpose for the purposes of evidencing their claim, helping to avoid accidental injustice in a Court or Tribunal, or more often, a lower level decision maker (e.g. an employee of a local council deciding what accommodations or adaptations to install in someone's home). For instance, the employer using AI to analyse recruitment interview videos might have to explain which portions of the video contributed negatively to the applicant's overall score; if these coincided with the applicant stammering, that would help evidence the claim.

Thus, the transparency and explainability of AI systems is a vital tool in the context of helping to challenge disability discrimination going forwards, perhaps especially in respect of pre-existing discrimination that still pervades the day-to-day lives of many people with disabilities. These are especially important when developing systems that support evidence gathering in relation to disability, which is a (sadly) important part of the day-to-day lives of many PWD: indeed, many of these proceedings can be distressing processes that take over the lives of claimants for many months (Ryan 2019).

5.2.4. *Advancing AT design and inclusive design in respect of better decision-making tools about disability*

In respect of decision-making tools in respect of disability, we would further observe that the intersection of data protection law and equality law can more directly advance AT design and similarly, the universal design of systems also used by PWD. First, the range of information that can be made available (see 5.2.2. above) by way of the GDPR will likely ensure that practices and design decisions that are potentially discriminatory (including indirectly, or by way of failing to make reasonable adjustments) against PWD are more easily identified. Not only should this advance best practice, but it also has in important flipside, in that there would be less room for defence under Part 8 of the Equality Act (2010) (and similar provisions elsewhere) in respect of those who promote inappropriate (and potentially discriminatory) technology design practices in respect of PWD. As such, better data is not only a lever for improving our knowledge of AT best practice, but also over time to ensure that only qualified decision makers inform the design of systems that are used by PWD (whether as AT, or from a universal design perspective).

---

[112] See https://www.stammeringlaw.org.uk/employment/recruitment-promotion/discrimination-computer-algorithm-recruitment/#evidence (archived at https://perma.cc/2P6P-QMJC)

Second, it is possible that in some contexts, the use of AI systems could help alleviate or avoid the prejudices of human decision-makers with regard to PWD, and depending on how they handle disability data disclosure, might alleviate some concerns of PWD about the disclosure of their disability to a human decision-maker. As Guo et al 2019 argue, with the right 'design guidelines, datasets, algorithmic techniques, and error metrics', AI systems have 'enormous potential to benefit PWD' (Guo et al 2019). The development of algorithmic hiring tools could also force e.g. employers to re-think their existing, possibly discriminatory evaluation criteria (e.g. traits like 'optimism' or 'intensity' (Centre for Democracy & Technology, 2020)). Despite the various problems AI may present for PWD noted above, it is also worth observing the scale of discrimination (with some groups of disabled people having - for example - very few employment prospects due to pervasive discrimination) against PWD means that use of AI systems could in some contexts be the lesser evil, especially given the asymmetrical nature of disability protection removing the possibility of direct discrimination claims being brought by people without disabilities (see 5.1 above). Furthermore, the use of automated AI systems could potentially limit the risk of further disclosures that often results from careless indiscretion on the part of human decision-makers. As such, we can see that some PWD (although certainly not all) might prefer *fair* automated decision making on job applications and other decisions, for example in determining who is invited to interview. However, the reality is that at present, both human and AI-assisted decision-making processes are rife with inequities for PWD, and this kind of 'lesser evil' argument is often used in an attempt to justify what should be deemed an unacceptably low bar.

Finally, we observe that the GDPR and DP law, *if* used appropriately, could greatly increase the availability and quality of testing and training data for AI systems designed to support PWD and to further their interests. After all, data protection may also encourage the collection of more granular data needed to ensure that AI systems based on machine learning work accurately for PWD. In turn, that may mean more trust in these systems, if they become increasingly accurate (and more people in turn would support them – serving as a form of virtuous circle). We argue that ensuring appropriate sensitivity (and confidentiality) and the promotion of trust in how disability-specific information will be processed should be seen as an important goal in the design of AI based AT, especially that which requires the collection of relatively large datasets to support it. Similarly, data protection law is a versatile toolkit, yet is also the only framework that can realistically enable this: as such, a detailed consideration of the more subtle aspects of this area of law should be seen to be an important design consideration going forward.

5.2.5. *Advancing the role of Disability Rights Organisations in AI Fairness*

Some of the protections in DP law place a particular emphasis upon independent non-profit organisations having a mediating role. In particular, one of the substantial public interest conditions laid out in UK data protection law gives a particular role in the processing of special category data to 'non-profit' organisations which "*provide support to individuals with a particular disability or medical condition*"[113]: thus providing this group of organisations a key potential role in furthering AI fairness and allowing them to process data in a way that other organisations cannot. We argue that this presents an important opportunity for user-led disability rights organisations (i.e. disabled people's organisations) to be those organisations that serve this mediating role, and thus be at the forefront of ensuring a new generation of fair AT and AI related best practices. This is especially so when set against the history of disability rights, with the rallying slogan of '*Nothing about us without us*' having been at the forefront of disability rights advocacy (Scotch 2009; Charlton 2009). The reason for this approach is in part an unhappy history of institutionalisation, where many PWD have had greatly reduced liberties and were often (and some remain) shut off from society or segregated from it. Indeed, this has perhaps come full circle with the advance of DP law, given that one of the relatively historic objections to institutionalisation being the lack of privacy that comes with it (TenBroek 1966; Lazar, Wentz, & Winckler 2017).

---

[113] Schedule 1, Paragraph 16 of the Data Protection Act (2018).

At the same time, there is a wider recognition that involvement of PWD in policy and the design of interactive systems ensures more appropriate outcomes for this group, and moreover, outcomes that are likely to be acceptable to them (with DIY / Bespoke AT being perhaps the defining example of this within the AT community (Hurst & Tobias 2011; Ellis et al 2020)). What's more, we also notice that respected user-led disabled persons organisations could serve as a trusted third party, who put into practice fair AI for PWD going forward: for instance, they could be involved on behalf of individuals in the context of Data Protection Impact Assessments (as discussed above). This would mean an expanded role for disabled people's organisations in the evaluation of AI systems in relation to disability and an opportunity to shape their development and deployment as a result. Indeed, noting the observations made above concerning class actions, disabled persons organisations could have a triple role as evaluators, policy shapers and where appropriate, using their trusted status to support class actions in relation to disability as a further form of advocacy in respect of these issues. Thus, DP combined with AI not only presents a wide range of new rights, but a genuine opportunity for PWD to influence their implementation on the ground.

6. **CONCLUSION**

In this article, we have examined the application of data protection law and equality law to the concern of AI fairness. We have found that a careful consideration of the intersection of these two areas of law allows for opportunities (and rights) that are distinctive to people with disabilities, thus necessitating a different approach towards 'AI Fairness' for this population at a fundamental level. While some prior work in AI fairness uses generic notions of 'bias' and 'fairness' which are subject to significant disagreement, and applies them to deal with different kinds of discrimination and injustice, we argue there are advantages to grounding AI fairness in the relatively well-defined and internationally agreed-upon concepts in disability discrimination law and data protection law (whilst recognizing their imperfections). At the same time, we have identified how a more expansive consideration of AI Fairness can provide new opportunities for PWD and the development of assistive technologies, positively advancing disability rights. Overall, we hope that the assistive technology community can take advantage of the legislative frameworks that constrain potentially unfair AI, and promote the development and availability of AI systems that better reflect universal design principles going forward.

Mittler, Peter. "The UN Convention on the Rights of Persons with Disabilities: Implementing a Paradigm Shift"; Journal of Policy and Practice in Intellectual Disabilities 12, no. 2 (2015): 79-89.

Morris, Meredith Ringel. "AI and Accessibility: A Discussion of Ethical Considerations." arXiv preprint arXiv:1908.08939 (2019).

Nakamura, Karen. "My Algorithms Have Determined You're Not Human: AI-ML, Reverse Turing-Tests, and the Disability Experience." The 21st International ACM SIGACCESS Conference on Computers and Accessibility. 2019.

Pedreshi, Dino, Ruggieri, Salvatore, and Turini, Franco. Discrimination-aware data mining. InProceedings of the 14th ACM SIGKDD international conference on Knowledge discovery anddata mining, pages 560–568. ACM, 2008

Scotch, Richard K. ""Nothing about us without us": Disability rights in America." OAH Magazine of History 23.3 (2009): 17-22.

TenBroek, Jacobus. "The right to live in the world: The disabled in the law of torts." Calif. L. Rev. 54 (1966): 841.

Trewin, Shari. "AI fairness for people with disabilities: Point of view." CACM

Trewin, Shari, et al. "Considerations for AI fairness for people with disabilities." AI Matters 5.3 (2019): 40-63.

Raghavan, Manish, et al. "Mitigating bias in algorithmic hiring: Evaluating claims and practices." Proceedings of the 2020 Conference on Fairness, Accountability, and Transparency. 2020.

Ryan, Frances. Crippled. Verso Books, 2019.

Madnani, N., Loukina, A., Von Davier, A., Burstein, J., & Cahill, A. (2017, April). Building better open-source tools to support fairness in automated scoring. In Proceedings of the First ACL Workshop on Ethics in Natural Language Processing (pp. 41-52).

Nachbar, Thomas. "Algorithmic Fairness, Algorithmic Discrimination." Virginia Public Law and Legal Theory Research Paper 2020-11 (2020).

Wachter, Sandra, and Brent Mittelstadt. "A right to reasonable inferences: Re-thinking data protection law in the age of big data and AI." Colum. Bus. L. Rev. (2019): 494.

Whittaker, Meredith, et al. "Disability, Bias, and AI." AI Now Institute, November (2019).

Watson, Colin, Reuben Kirkham, and Ahmed Kharrufa. "PIP Kit: An Exploratory Investigation into using Lifelogging to support Disability Benefit Claimants." Proceedings of the 2020 CHI Conference on Human Factors in Computing Systems. 2020.
PREPRINT – 09.06.202136